**Magomet Yandiev**
Associate Professor, Department of Economics,
Lomonosov Moscow State University
mag2097@mail.ru

**Alexander Pakhalov**,
PG student, Department of Economics,
Lomonosov Moscow State University
pahalov@gmail.com

# The relationship between stock market parameters and interbank lending market: an empirical evidence

*Abstract*: The article presents calculations that prove practical importance of the earlier derived theoretical relationship between the interest rate on the interbank credit market, volume of investment and the quantity of securities tradable on the stock exchange.

## 1. Purpose

The paper "The Damped Fluctuations as a Base of Market Quotations" (M. Yandiev, 2011) theoretically substantiates the formula which shows the relationship between the interest rate on the interbank market, volume of investments on the stock market and quantity of securities tradable on the stock exchange. However, the paper did not contain calculations that prove applicability of the formula.

To continue, the purpose of this paper is to make calculations and to provide arguments that will prove or disprove the significance of the theoretical formula.

## 2. Brief description of the theoretical relationship

The starting point in generation of the formula comprises a number of assumptions which simplify understanding of the pricing process on the stock exchange. Major assumptions are as follows:

- out of all financial markets the stock market is the only one that exists;
- the market trades ordinary stocks only of a single issuer;
- no information is channeled to the market;
- no applications are filed on the market from clients of brokers.

In view of the motivation of the remaining market dealers, namely increase in company's equity price, we deal with the situation where there are no grounds to change stock prices on the exchange except used for speculations.

Then a number of assumptions become weaker. In particular, we admit existence of the interbank credit market of which attributes, including the interest rate, appear to be attractive for dealers to make investments, or invest alternatively on the stock market.

We assume that a trader is always prepared to admit some loss closing deals that seem to be unprofitable for him/her in order to wait for an appropriate moment and finally win. Hence, under the circumstances a dealer does not care about a loss equivalent to a part of his/her equity which may easily be recovered using an alternative source, i.e. interbank credit market. Then the largest daily limit to such loss may be found from the formula:

$$L = I * R * \frac{1}{365} \qquad [1]$$

- **L** is a loss or amount of funds that a dealer is prepared to lose when trading with a view to gain per day;
- **I** is the volume of speculative investments (amount of money on accounts in the authorized bank to the stock exchange and intended for speculations);
- **R** is the interest rate on one-day loans on the interbank credit market, in fractions;

Then the paper shows that inflow/outflow of trading resources to the stock exchange has an impact on the growth or reduction only in the number of deals closed on the market and that for a dealer the benchmark is always some mean loss per a single deal. Then the relationship between inflow/outflow of speculative investments and the number is as follows:

$$L = U * u \qquad [2]$$

- **U** is the total amount of stocks involved in deals;
- **u** is the mean loss per a deal involving one stock.

Further formulas 1 and 2 are equated on the basis of **L** parameter (loss):

$$u = I * R * \frac{1}{365} * \frac{1}{U} \qquad [3]$$

The **u** parameter is shifted to the right side since this is a constant and as such the formula demonstrates better the logic of market relationships. For instance, the greater speculative capital and interest rate on the interbank credit market the greater the volume of trading on the exchange.

### 3. Description of the data

To verify applicability of the formula, we use 2012 daily data provided by the Moscow Exchange[1], where at the moment 100% of money and financial assets were pre-deposited:

- Total amount of money deposited within the exchange system in m. rubles (analogue of **I** parameter, refer to Appendix 1).

  Number of stocks (blue chips) deposited in the clearing exchange system, in pcs (**U** parameter, refer to Appendix 2). We used data on 11 most liquid stocks rather than on all of them, i.e. blue chips: Sberbank (ordinary stocks and preference stocks), Gazprom ordinary stocks, GMKN Norilski Nickel ordinary stocks, VTB ordinary stocks, LUKoil ordinary stocks, Transneft preference stocks, Rosneft ordinary stocks, Rushydro ordinary stocks, Severstal ordinary stocks, FGC UES ordinary stocks.

[1] We thank Andrei Shemetov, Deputy Chairman of the Executive Board, and Alexander Schliappo, Managing Director for Moscow Exchange OJSC Process Systems Development, for assistance in acquisition of data required which are not publicly accessible.

- Fraction of blue chips in the total volume of stock trading, % (this information is needed to be sure that blue chips data is representative and reflect the situation on the stock market, refer to Appendix 3).

Additional data were obtained from the official site of the Bank of Russia. This data include one-day loan interest rates (see Appendix 4). In calculation of the **u** parameter we used data on each trading day and each out of 255 working days in 2012.

To verify applicability of the formula, we used two different approaches. In both approaches **U** parameter was taken both as quantity of all deposited stocks within the exchange system and as the quantity of securities involved in the stock exchange deals.

## 4. First approach. Formula verification based on standard deviation of the “u” variable

In the first approach the target was to make sure that **u** parameter is a relatively constant value, i.e. its standard variations are insignificant. The calculations have shown that the standard variations are below one hundredth of the mean price of a single stock. It allows us to recognize **u** parameter as a generally constant value (refer to Appendix 5). In addition, visual examination of the daily **u** parameter value has shown that it is slightly volatile (refer to Appendixes 6-7).

It is notable that in May 2012 the mean **u** parameter changed. It has grown about twice as much (we compare the mean **u** parameter from the beginning of the year to May 10 with the mean value from May 10 to the end of the year). Growth has been found in both variants of calculations. The change coincided in time with rapid amount of money deposited in the stock exchange trading system (refer to Appendix 1). It may be accounted for by the additional inflow of investments to the market which caused the increase in the risk the participants are prepared to take in closing the stock trading deals which gave impetus to **u** parameter growth.

In the same way we may account for the fact that **u** parameter unexpectedly grows on some days, in particular, on the eve of holidays: approach of the time when the market has a rest sharpens the participant's sense of uncertainty and risk of negative variations of quotations which reduces risk perception and causes **u** parameter growth.

Thus, volumes of funds and assets deposited in the system appreciably outweighed current needs for trade operations. For instance, out of 100 stocks deposited in the system the trade operations involved on average ten stocks while per 1 ruble of the market price of stocks also deposited in the system 65 kopecks were also deposited in the system (refer to Appendix 8). The fact evidences a super high protection against risks ensured on the Moscow Stock Exchange, i.e. 100% reservation of funds and assets. This pattern, however, significantly restricted the choice of participants, and switch to more liberal rules of funds and assets reservation to take place in 2013 will intensify activities on the financial markets without detriment to confidence.

## 5. Second approach. Formula verification based on linear regression

The second approach uses regression analysis of time series in order to identify relationship between variables of our model. The aim of this analysis is to verify relationship expected in the theoretical model (specifically in the formula [3]). Input time series for each of the six variables (see Appendix 9) consist of 255 daily observations. All calculations were made in EViews 7.0.

At the first stage of econometric analysis, we have tested all variables for unit roots. It is required that the various variables are stationary, because major part of econometric methodology is built upon the assumption of stationarity (Verbeek, 2004, p. 309-310). We have used an augmented Dickey–Fuller test (ADF) which is one of the most popular tests for a unit root in a time series sample. Lag length in each case was set based on the Schwarz information criterion (SIC). The results of unit root testing procedure can be found in Appendix 10.

ADF test has shown that some of variables are non-stationary. The use of non-stationary variables in linear regressions may result in invalid estimators. An important exception arises when these variables are cointegrated. In this case, the OLS estimator can give super consistent estimates of parameters (Verbeek, 2004, p. 314-315). In our case, residuals of two regressions based on two different ways of **"u"** calculations are stationary at the 1% level of significance. Thus, variables in both cases are cointegrated. This allows us to make a number of conclusions based on linear regressions provided in Appendix 11.

Both equations are significant, and the relationship between parameter **"u"** and other variables corresponds to the theoretical formula. Variables **R** and **I** have positive coefficients in the equations (it means the direct relationship with the dependent variable **u**), and the variable **U** has a negative coefficient (it means the inverse relationship with the dependent variable **u**). This conclusion has the same significance for the equations based on two methods of **"u"** calculation.

## 6. Summary

Calculations made in the first and second approaches have shown that the formula, in general, correctly reflected the relationship between parameters of the interbank credit market and Moscow stock exchange market in 2012 which is indicative of applicability of the formula.

We recommend to apply the formula in financial markets regulation. For instance, it is possible to use it in prediction of effect of critical situations on the markets in the event of appreciable inflow-outflow of money and securities and rapid change in credit interest rates on the interbank credit markets.

## 7. References

- Yandiev, Magomet. The Damped Fluctuations as a Base of Market Quotations. Economics and Management, № 16, 2011. ISSN 1822-6515. URL: http://ssrn.com/abstract=1919652
- Verbeek, Marno. A guide to modern econometrics. 2nd edition. – Chichester: John Wiley & Sons Ltd, 2004. ISBN 0-470-85773-0
- Яндиев М. Концепция фондового кредита в структуре рыночной котировки. Рынок ценных бумаг, № 6, 2011 г.

Appendix 1.

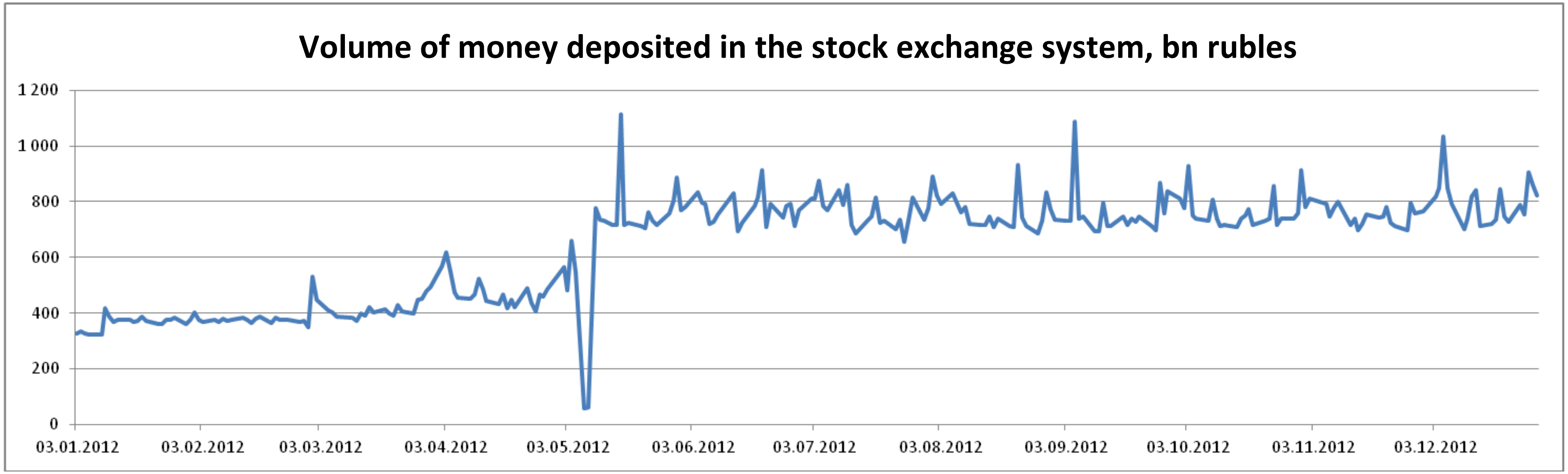

Appendix 2. **Quantity of stocks (blue chips) deposited in the clearing system of the exchange**

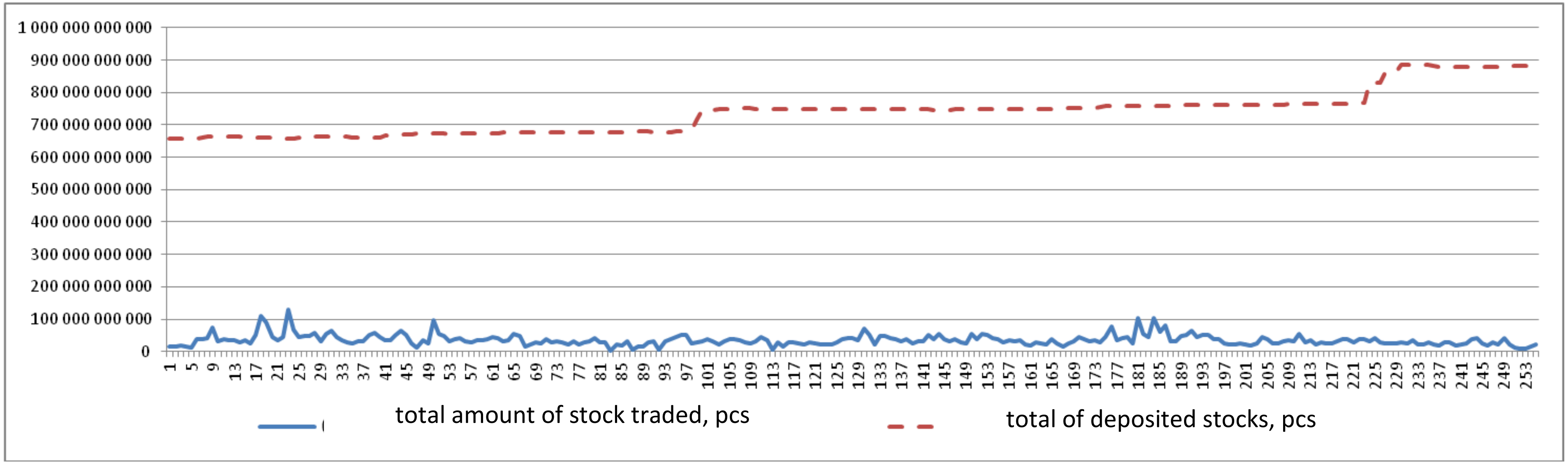

Appendix 3.

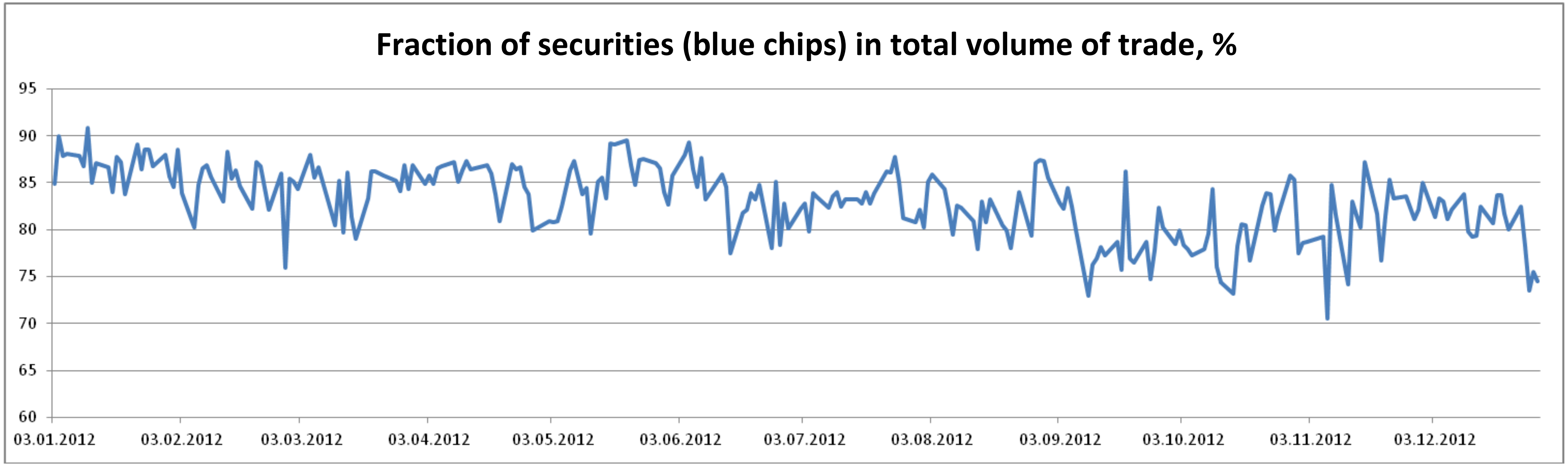

Appendix 4.

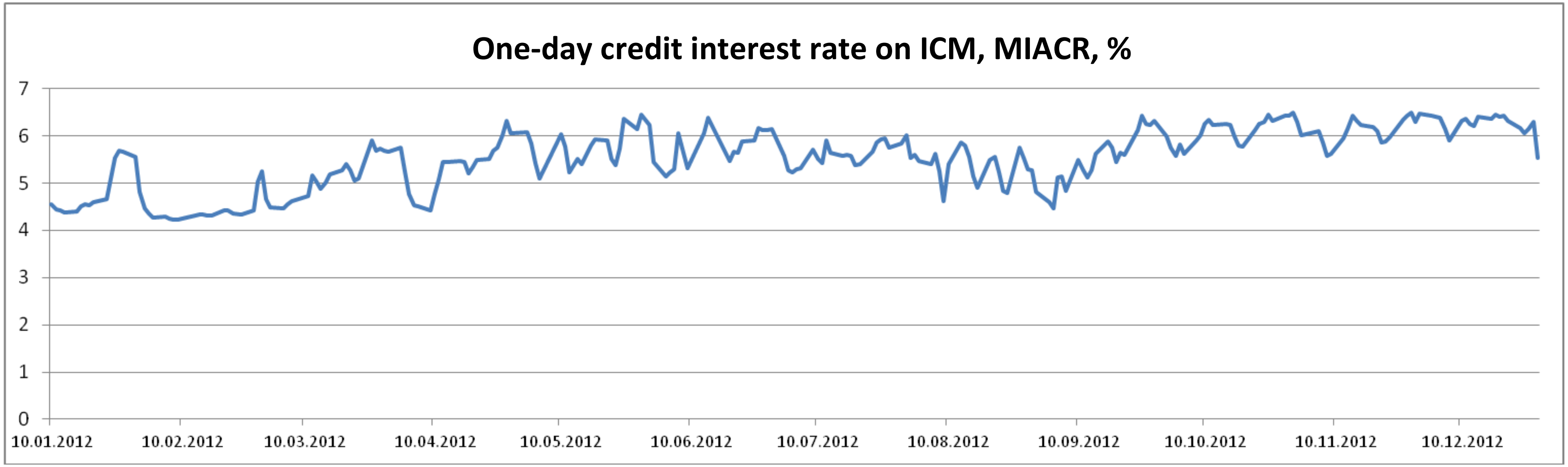

Appendix 5.

**u parameter calcualtions**

| "u" parameter, kopecks, in 2012 | Calculation, where U parameter is quantity of stocks involved in trade | Calculation, where U parameter is total amount of deposited stocks |
|---|---|---|
| Arithmetic mean, kopecks | 51 | 1.9 |
| Volatility, kopecks | 40 | 0.6 |
| Mean price of one stock, rubles | 5,320 | |

Appendix 6.

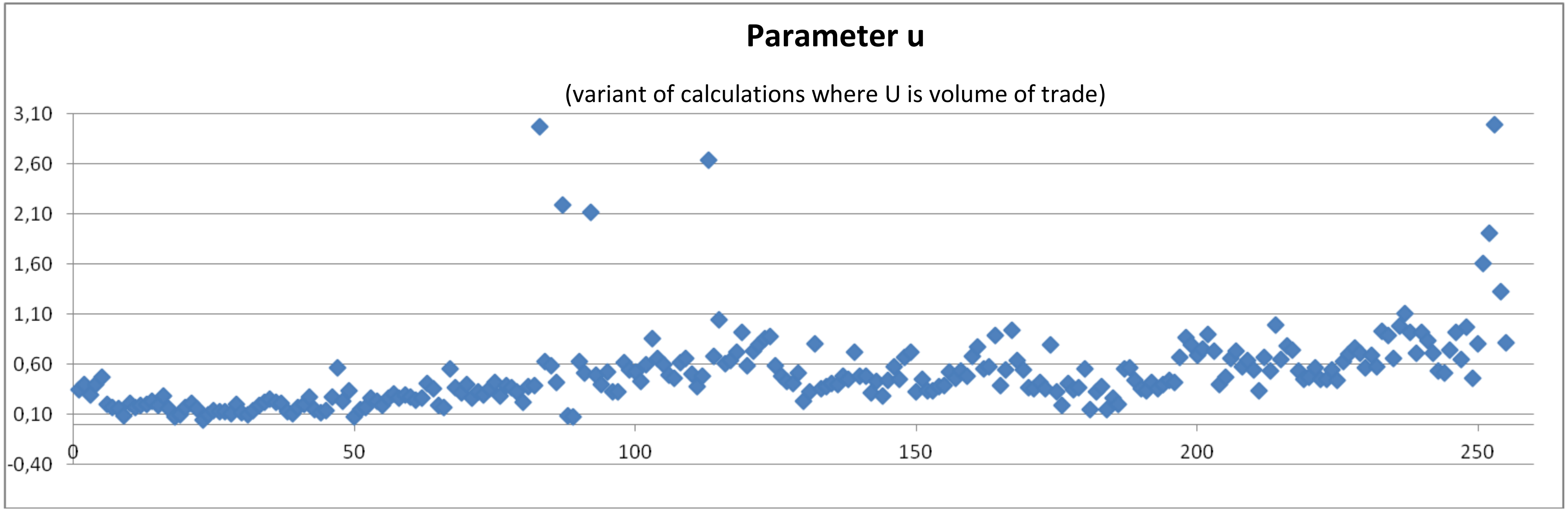

Appendix 7.

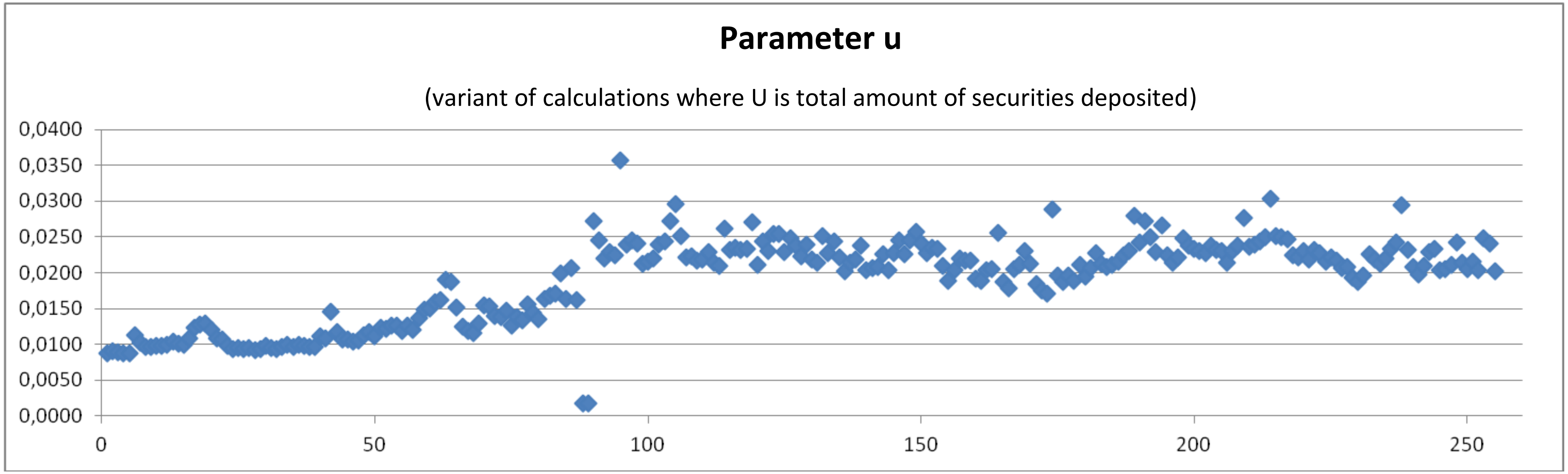

Appendix 8.

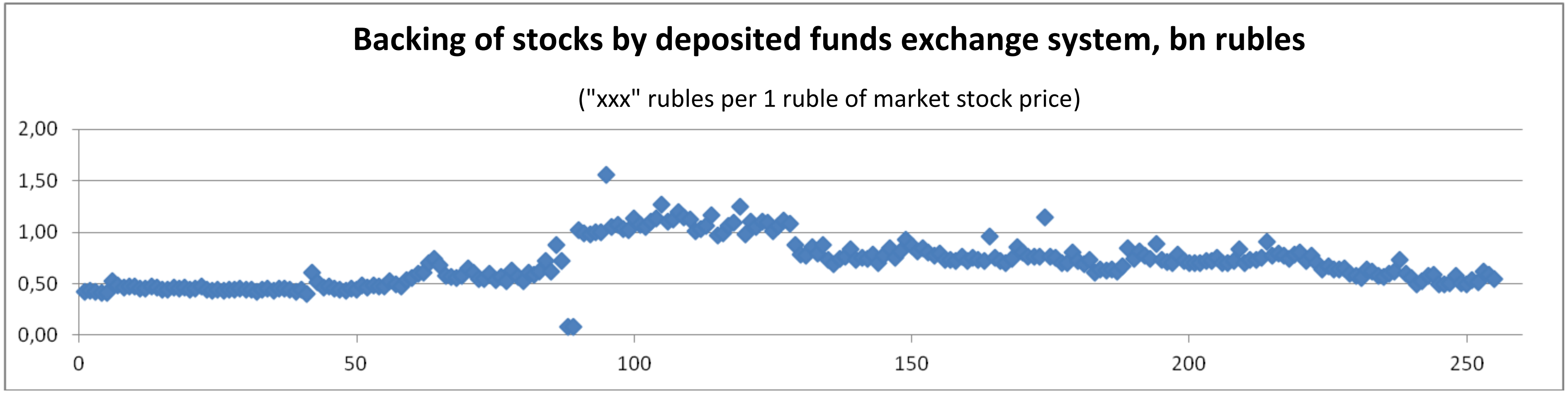

Appendix 9. List of variables with their definitions

| **Variable name in theoretical model** | **Variable name in EViews tables** | **Definition** |
|---|---|---|
| u | U_SMALL_VOL | Mean loss per a deal involving one stock (calculated using the amount of stocks involved in deals) |
| u | U_SMALL_DEP | Mean loss per a deal involving one stock (calculated using the amount of deposited stocks) |
| I | I | Volume of speculative investment |
| R | R | One-day loan interest rate on the interbank lending market |
| U | U_BIG_VOL | Total amount of stocks involved in the stock exchange deals |
| U | U_BIG_DEP | Total amount of deposited stocks within the exchange system |

Appendix 10. Unit root testing

10.1) Unit root test for "U_SMALL_VOL"

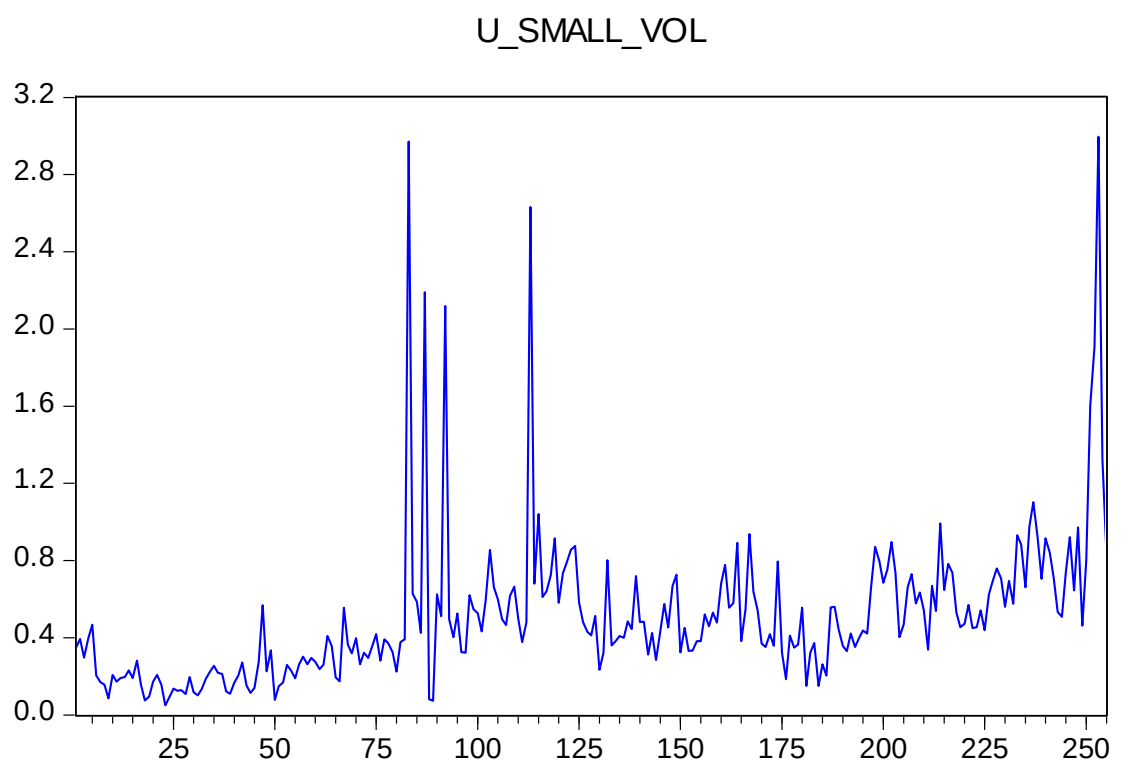

ADF test results (level):

Null Hypothesis: U_SMALL_VOL has a unit root
Exogenous: Constant
Lag Length: 3 (Automatic - based on SIC, maxlag=5)

| | | t-Statistic | Prob.* |
|---|---|---|---|
| Augmented Dickey-Fuller test statistic | | -3.370369 | 0.0129 |
| Test critical values: | 1% level | -3.456302 | |
| | 5% level | -2.872857 | |
| | 10% level | -2.572875 | |

*MacKinnon (1996) one-sided p-values.

## 10.2) Unit root test for “U_SMALL_DEP”

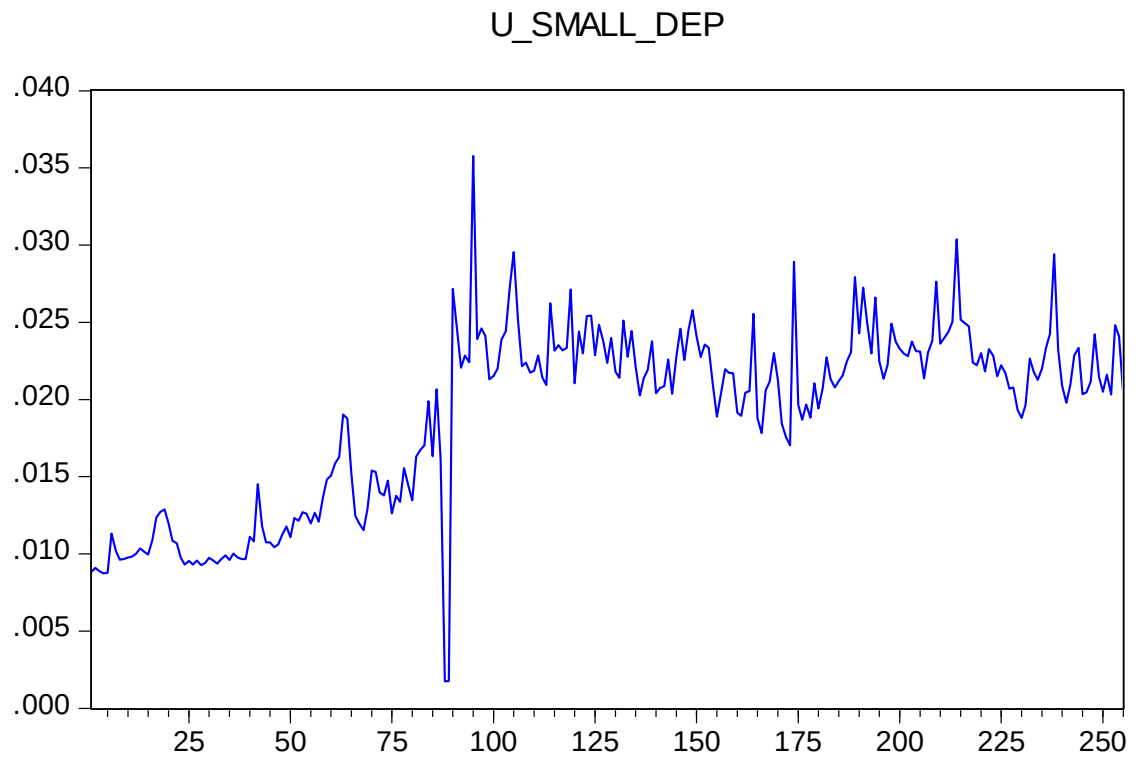

### ADF test results (level):

Null Hypothesis: U_SMALL_DEP has a unit root
Exogenous: Constant
Lag Length: 4 (Automatic - based on SIC, maxlag=5)

| | | t-Statistic | Prob.* |
|---|---|---|---|
| Augmented Dickey-Fuller test statistic | | -2.152787 | 0.2244 |
| Test critical values: | 1% level | -3.456408 | |
| | 5% level | -2.872904 | |
| | 10% level | -2.572900 | |

*MacKinnon (1996) one-sided p-values.

### ADF test results (first differences):

Null Hypothesis: D(U_SMALL_DEP) has a unit root
Exogenous: Constant
Lag Length: 3 (Automatic - based on SIC, maxlag=5)

| | | t-Statistic | Prob.* |
|---|---|---|---|
| Augmented Dickey-Fuller test statistic | | -13.40007 | 0.0000 |
| Test critical values: | 1% level | -3.456408 | |
| | 5% level | -2.872904 | |
| | 10% level | -2.572900 | |

*MacKinnon (1996) one-sided p-values.

## 10.3) Unit root test for “I"

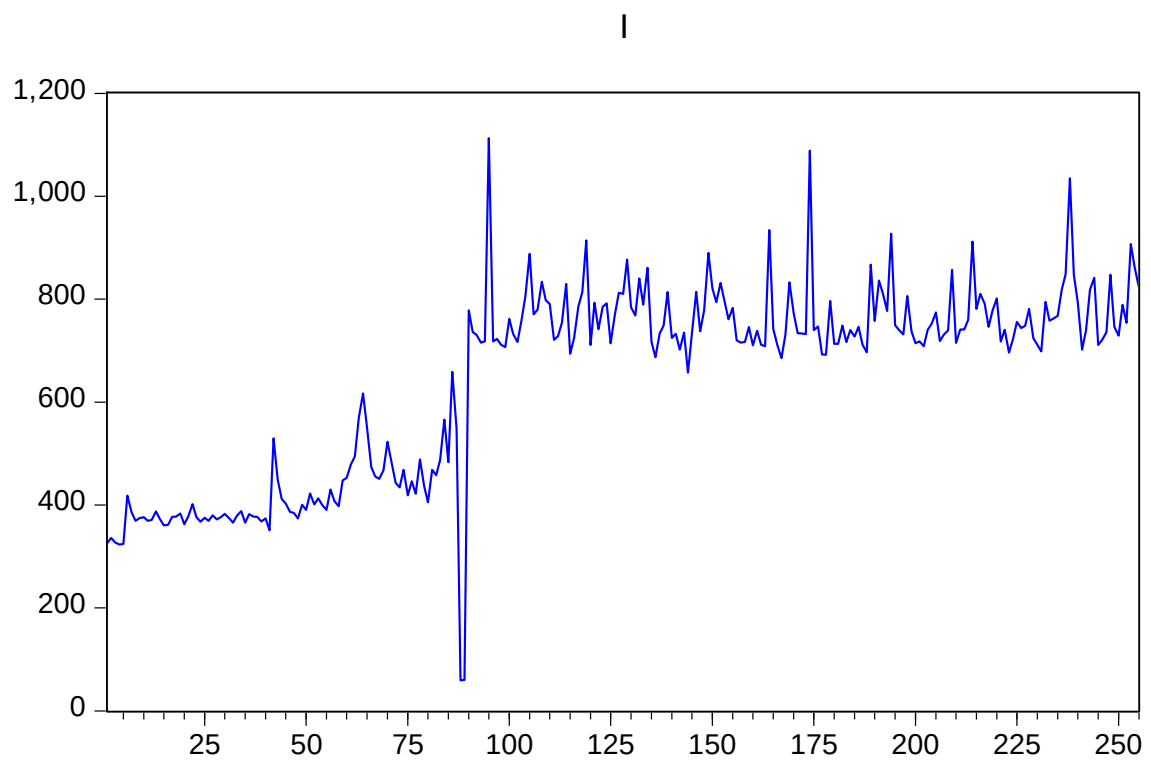

ADF test results (level):

Null Hypothesis: I has a unit root
Exogenous: Constant
Lag Length: 4 (Automatic - based on SIC, maxlag=5)

| | | t-Statistic | Prob.* |
|---|---|---|---|
| Augmented Dickey-Fuller test statistic | | -1.801486 | 0.3793 |
| Test critical values: | 1% level | -3.456408 | |
| | 5% level | -2.872904 | |
| | 10% level | -2.572900 | |

*MacKinnon (1996) one-sided p-values.

ADF test results (first differences):

Null Hypothesis: D(I) has a unit root
Exogenous: Constant
Lag Length: 3 (Automatic - based on SIC, maxlag=5)

| | | t-Statistic | Prob.* |
|---|---|---|---|
| Augmented Dickey-Fuller test statistic | | -14.21234 | 0.0000 |
| Test critical values: | 1% level | -3.456408 | |
| | 5% level | -2.872904 | |
| | 10% level | -2.572900 | |

*MacKinnon (1996) one-sided p-values.

## 10.4) Unit root test for "R"

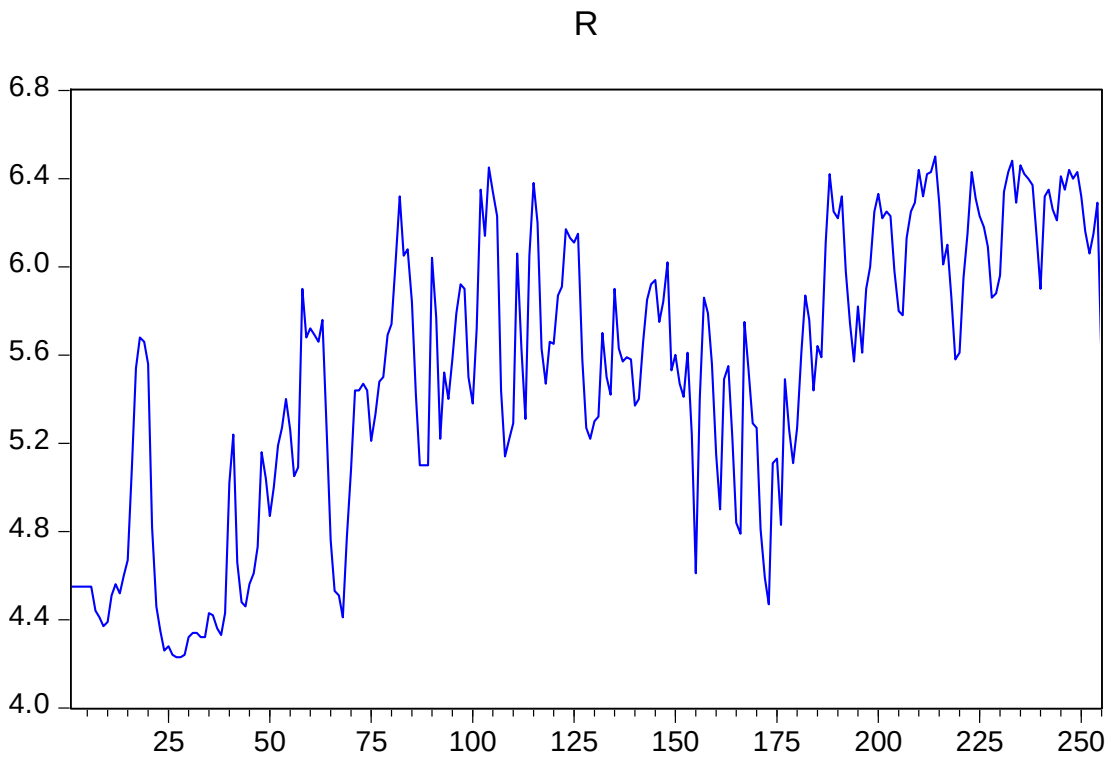

ADF test results (level):

Null Hypothesis: R has a unit root
Exogenous: Constant
Lag Length: 2 (Automatic - based on SIC, maxlag=5)

| | | t-Statistic | Prob.* |
|---|---|---|---|
| Augmented Dickey-Fuller test statistic | | -3.349603 | 0.0138 |
| Test critical values: | 1% level | -3.456197 | |
| | 5% level | -2.872811 | |
| | 10% level | -2.572851 | |

*MacKinnon (1996) one-sided p-values.

## 10.5) Unit root test for "U_BIG_VOL"

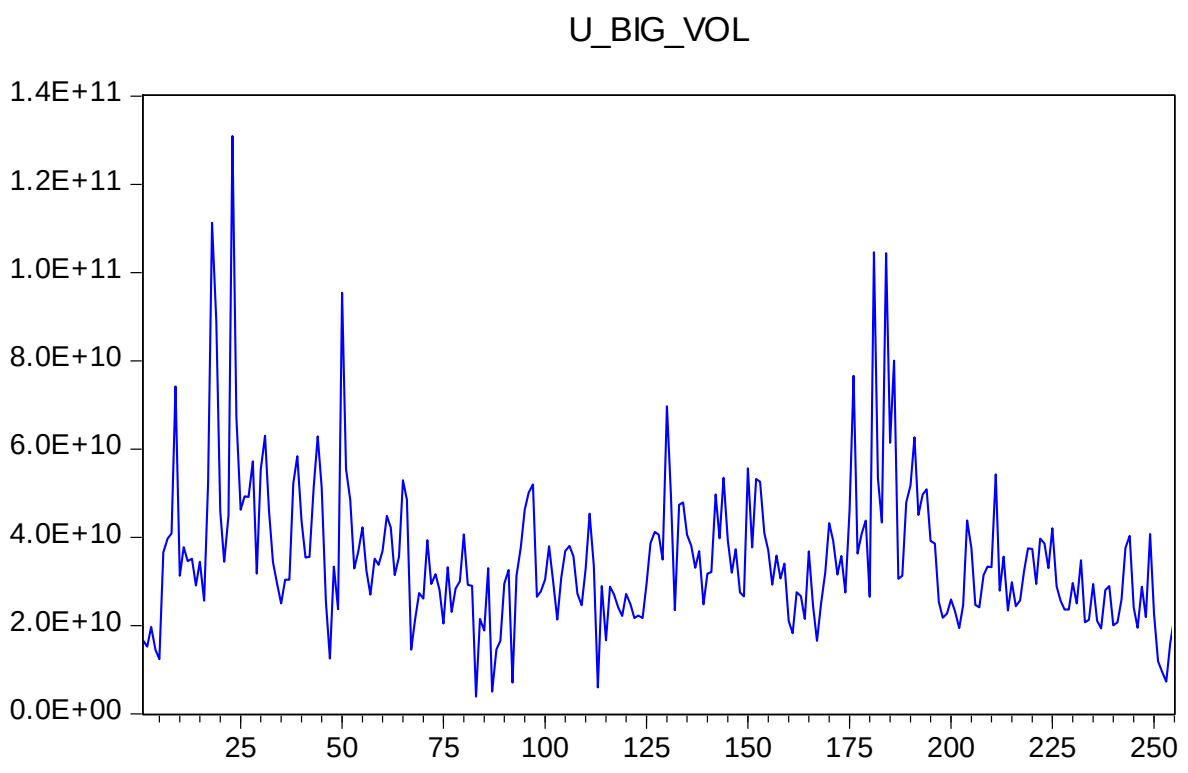

## ADF test results (level):

Null Hypothesis: U_BIG_VOL has a unit root
Exogenous: Constant
Lag Length: 4 (Automatic - based on SIC, maxlag=5)

| | | t-Statistic | Prob.* |
|---|---|---|---|
| Augmented Dickey-Fuller test statistic | | -3.681785 | 0.0049 |
| Test critical values: | 1% level | -3.456408 | |
| | 5% level | -2.872904 | |
| | 10% level | -2.572900 | |

*MacKinnon (1996) one-sided p-values.

## 10.6) Unit root test for "U_BIG_DEP"

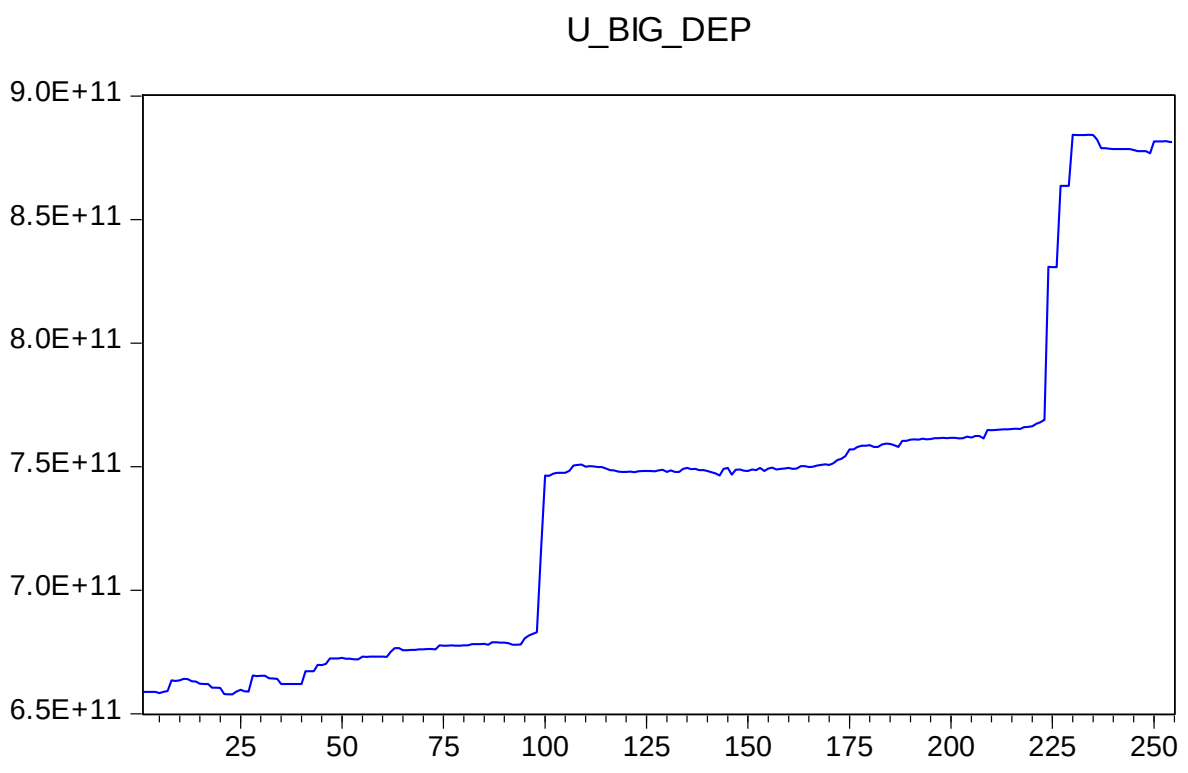

## ADF test results (level):

Null Hypothesis: U_BIG_DEP has a unit root
Exogenous: Constant
Lag Length: 3 (Automatic - based on SIC, maxlag=5)

| | | t-Statistic | Prob.* |
|---|---|---|---|
| Augmented Dickey-Fuller test statistic | | -0.331998 | 0.9168 |
| Test critical values: | 1% level | -3.456302 | |
| | 5% level | -2.872857 | |
| | 10% level | -2.572875 | |

*MacKinnon (1996) one-sided p-values.

ADF test results (first differences):

Null Hypothesis: D(U_BIG_DEP) has a unit root
Exogenous: Constant
Lag Length: 2 (Automatic - based on SIC, maxlag=5)

| | | t-Statistic | Prob.* |
|---|---|---|---|
| Augmented Dickey-Fuller test statistic | | -5.954171 | 0.0000 |
| Test critical values: | 1% level | -3.456302 | |
| | 5% level | -2.872857 | |
| | 10% level | -2.572875 | |

*MacKinnon (1996) one-sided p-values.

ADF test results – summary:

| Variable name in EViews tables | ADF test results |
|---|---|
| u_small_vol | Variable is stationary at the 5% level of significance |
| u_small_dep | Variable is stationary in first differences at the 1% level of significance |
| i | Variable is stationary in first differences at the 1% level of significance |
| r | Variable is stationary at the 5% level of significance |
| u_big_vol | Variable is stationary at the 1% level of significance |
| u_big_dep | Variable is stationary in first differences at the 1% level of significance |

Appendix 11. Testing for cointegration and linear regressions

11.1) Regression 1: U is the quantity of securities involved in the stock exchange deals

Linear regression:

Dependent Variable: U_SMALL_VOL
Method: Least Squares
Sample: 1 255
Included observations: 255

| Variable | Coefficient | Std. Error | t-Statistic | Prob. |
|---|---|---|---|---|
| C | 0.078587 | 0.179113 | 0.438757 | 0.6612 |
| U_BIG_VOL | -1.19E-11 | 1.10E-12 | -10.84576 | **0.0000** |
| R | 0.079999 | 0.036926 | 2.166459 | **0.0312** |
| I | 0.000639 | 0.000124 | 5.133514 | **0.0000** |

| | | | |
|---|---|---|---|
| R-squared | **0.480387** | Mean dependent var | 0.508588 |
| Adjusted R-squared | **0.474177** | S.D. dependent var | 0.397053 |
| S.E. of regression | 0.287918 | Akaike info criterion | 0.363279 |
| Sum squared resid | 20.80708 | Schwarz criterion | 0.418829 |
| Log likelihood | -42.31813 | Hannan-Quinn criter. | 0.385624 |
| F-statistic | 77.35073 | Durbin-Watson stat | 1.676819 |
| Prob(F-statistic) | **0.000000** | | |

ADF test[2] results for residuals:

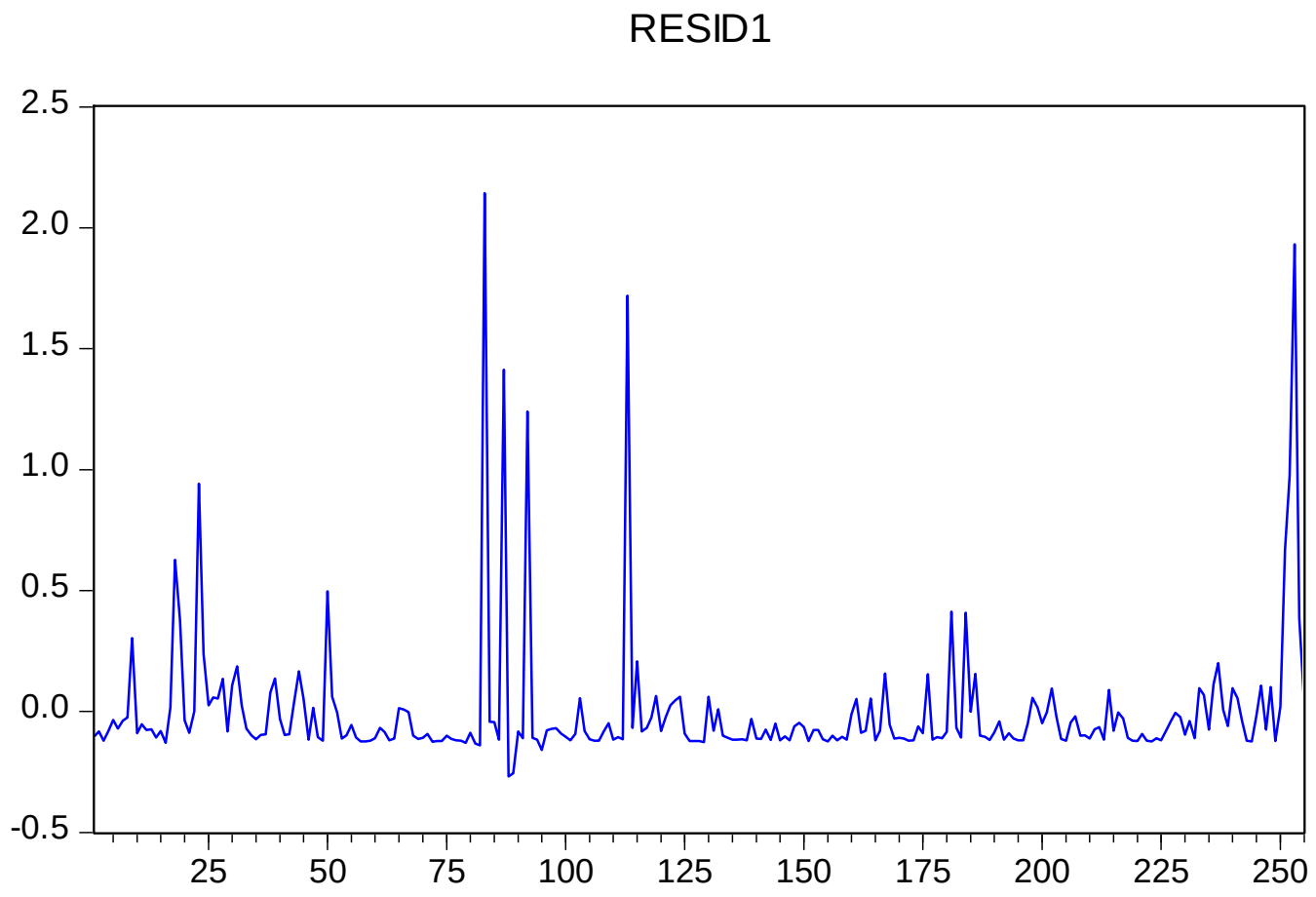

Null Hypothesis: RESID1 has a unit root
Exogenous: Constant
Lag Length: 0 (Automatic - based on SIC, maxlag=5)

| | | t-Statistic | Prob.* |
|---|---|---|---|
| Augmented Dickey-Fuller test statistic | | -13.49351 | 0.0000 |
| Test critical values: | 1% level | **-4.64** | |
| | 5% level | **-4.10** | |
| | 10% level | **-3.81** | |

11.2) Regression 2: U is the quantity of all deposited stocks within the exchange system

Linear regression:

Dependent Variable: U_SMALL_DEP
Method: Least Squares
Sample: 1 255
Included observations: 255

| Variable | Coefficient | Std. Error | t-Statistic | Prob. |
|---|---|---|---|---|
| C | 0.001134 | 0.000319 | 3.558902 | 0.0004 |
| U_BIG_DEP | -2.61E-14 | 5.70E-16 | -45.86139 | **0.0000** |
| R | 0.003287 | 5.24E-05 | 62.76459 | **0.0000** |
| I | 2.95E-05 | 1.89E-07 | 155.4837 | **0.0000** |

| | | | |
|---|---|---|---|
| R-squared | **0.995887** | Mean dependent var | 0.018954 |
| Adjusted R-squared | **0.995838** | S.D. dependent var | 0.005846 |
| S.E. of regression | 0.000377 | Akaike info criterion | -12.91226 |
| Sum squared resid | 3.57E-05 | Schwarz criterion | -12.85671 |
| Log likelihood | 1650.314 | Hannan-Quinn criter. | -12.88992 |
| F-statistic | 20259.96 | Durbin-Watson stat | 0.531755 |
| Prob(F-statistic) | **0.000000** | | |

ADF test[3] results for residuals:

[2] Here we use asymptotic critical values residual unit root tests for cointegration with constant term (Davidson and MacKinnon, 1993). See, e.g.: Verbeek, 2004, p. 316

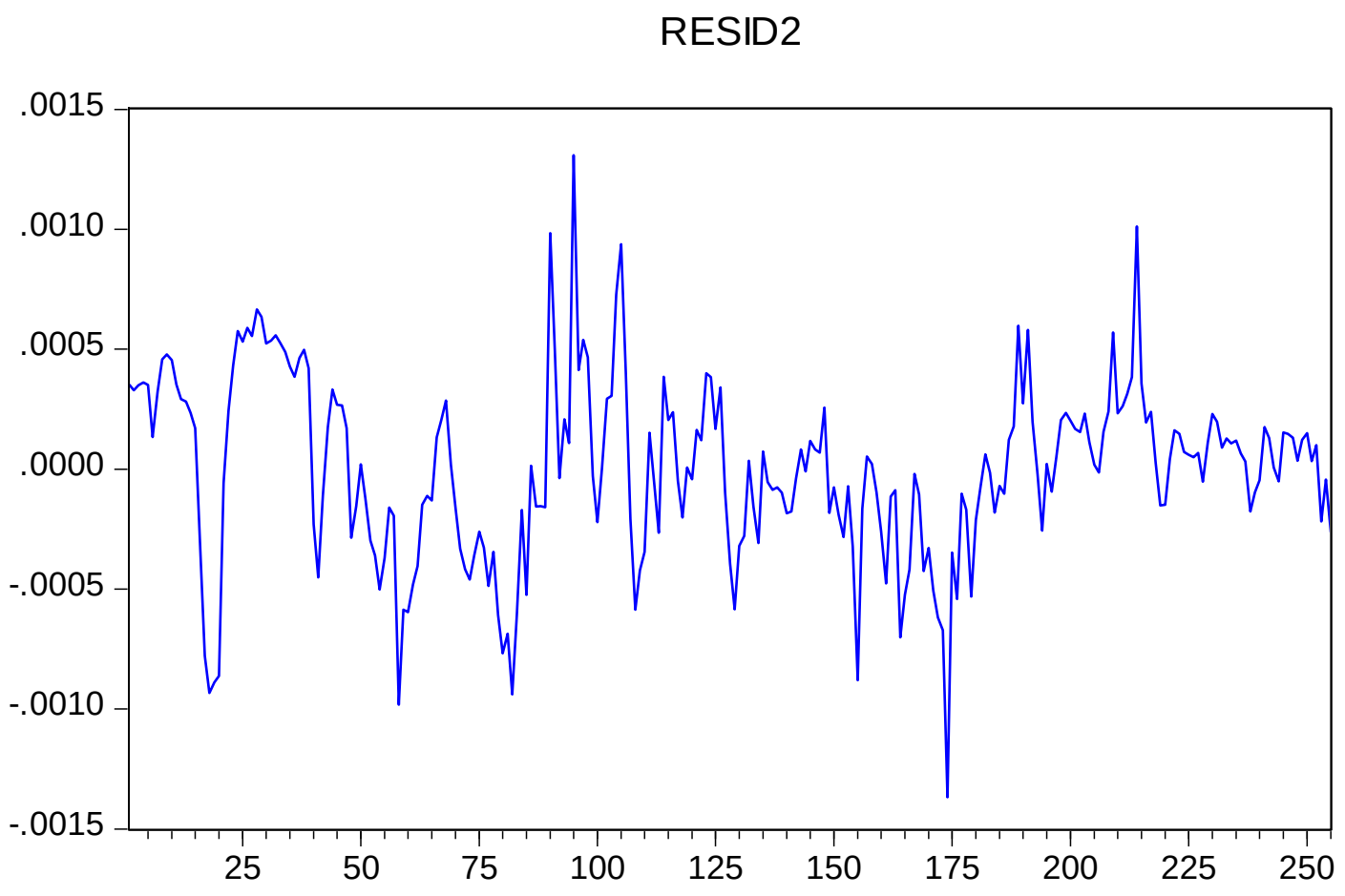

Null Hypothesis: RESID2 has a unit root
Exogenous: Constant
Lag Length: 0 (Automatic - based on SIC, maxlag=5)

| | | t-Statistic | Prob.* |
|---|---|---|---|
| Augmented Dickey-Fuller test statistic | | -6.243297 | 0.0000 |
| Test critical values: | 1% level | **-4.64** | |
| | 5% level | **-4.10** | |
| | 10% level | **-3.81** | |

*MacKinnon (1996) one-sided p-values.

[3] Here we use asymptotic critical values residual unit root tests for cointegration with constant term (Davidson and MacKinnon, 1993). See, e.g.: Verbeek, 2004, p. 316